\documentclass[aps,pra,twocolumn,groupedaddress,superscriptaddress,bibnotes,amsfonts,
citeautoscript,a4paper]{revtex4-1}
\usepackage{latexsym}
\usepackage{amsmath}
\usepackage{amssymb}
\usepackage{graphicx}
\usepackage[T1]{fontenc}
\usepackage[open]{bookmark}
\usepackage{hyperref}
\usepackage{subcaption}
\usepackage{graphicx}
\usepackage{epstopdf}

\hypersetup{colorlinks=true,allcolors=blue}

\begin{document}

\title{Linear and Nonlinear Characterization of broadband integrated Si-rich silicon nitride racetrack ring resonator for on-chip applications}

\author{Partha Mondal}
\altaffiliation[Present address: ]{ Department of Computer, Electrical and Mathematical Science and Engineering, King Abdullah University of Science and Technology (KAUST), Saudi Arabia}
\affiliation{Centre for Nano Science and Engineering (CeNSE), Indian Institute of Science, Bangalore, India}

\author{Venkatachalam P}
\affiliation{Centre for Nano Science and Engineering (CeNSE), Indian Institute of Science, Bangalore, India}
\author{Radhakant Singh}
\affiliation{Centre for Nano Science and Engineering (CeNSE), Indian Institute of Science, Bangalore, India}
\author{Sneha Shelwade}
\affiliation{Centre for Nano Science and Engineering (CeNSE), Indian Institute of Science, Bangalore, India}

\author{Gali Sushma}
\affiliation{Centre for Nano Science and Engineering (CeNSE), Indian Institute of Science, Bangalore, India}
\author{Shankar K Selvaraja}

\affiliation{Centre for Nano Science and Engineering (CeNSE), Indian Institute of Science, Bangalore, India}
\date{\today}

\begin{abstract}
We demonstrate the linear and nonlinear characterization of plasma-enhanced chemical vapor deposited silicon-rich silicon nitride (SRSN) racetrack ring resonator for on-chip application within the telecommunication wavelength range. The SRN waveguide parameters are optimized by employing the refractive index profile measured by ellipsiometry to achieve flat dispersion in the telecom band. Furthermore, we measure the thermo-optic coefficient (TOC) of the micro-resonator by analyzing the temperature-dependent transmission spectra and assessing it to be \(3.2825\) $\times$ \(10^{-5}\) \(^o{} C^{-1}\). Additionally, we perform power-dependent transmission spectra to investigate the effect of local heating and nonlinear absorption. The power-dependent transmission spectra exhibit a blue-shifting of the resonance peak in the visible and near-IR regions, which indicates the presence of nonlinear losses in that range. The power-dependent transmission spectra almost remain unchanged in the telecom band, revealing the absence of nonlinear losses and excellent thermal stability in that wavelength range. Our experimental results reveal that the SRSN-based structure can be employed potentially to realize linear and nonlinear applications in the telecom band.
\end{abstract}

\maketitle

\section{Introduction}

Over the years, Si-photonics has emerged as a promising CMOS-compatible material platform for fabricating low-cost, scalable integrated components for on-chip applications \cite{soref2006past,thomson2016roadmap}. Owing to the high-index contrast, high Kerr coefficient, transparency over a broad wavelength range (telecom to mid-IR), and enhanced device performance, Silicon-on-insulator (SOI) technologies in Si-photonics has been widely accepted for the future generation of CMOS integrated circuits (ICs). Researchers have  harnessed Si-photonics and employed in a plethora of applications, including high-speed data processing, sensing \cite{rogers2021universal,dorfner2008silicon}, nonlinear and quantum photonics \cite{harris2016large,leuthold2010nonlinear}, WDM systems \cite{fang2010wdm,moscoso20188}, and all-optical signal processing \cite{koos2007nonlinear}. However, despite the tremendous achievement of Si photonics, the small electronic band gap of Si (1.12 eV) \cite{leuthold2010nonlinear} imposes a fundamental limitation owing to the large two-photon (TPA) and free-carrier absorption (FCA) losses below 2.2 \(\mu\)m wavelength \cite{yin2007soliton}. These nonlinear losses are detrimental to the optical performance even at low power and prevent the widespread adoption of the Si platform in the telecommunication band \cite{yin2007impact,tsang2002optical,haldar2019free}. On the other hand, Si exhibits high thermo-optic coefficient (TOC) \((dn/dt = 1.86\) $\times$ \(10^{-4} K^{-1})\) which causes the SOI devices strongly thermally sensitive \cite{dong2010low}. The optical properties of Si-based devices can be significantly affected either by local heating causes due to the launching of high-power sources or changes in the environmental temperature. Therefore, Si based photonic devices are facing limitations in the application corresponding to high-temperature variation. Different approaches have been reported to reduce the thermal effect of Si-based devices, such as the incorporation of external metal heater for optimal heating \cite{manipatruni2008wide}, deposition of a cladding layer of Si with a material with negative TOC to compensate the positive TOC of Si \cite{han2007temperature,qiu2016athermal}, and so on. The limitation of the Si platform leads to exploring a new flexible CMOS-compatible platform carrying a refractive index lying between Si and \(SiO_{2}\), which can be realized for a multitude of photonics applications. In the quest for proper substitution of the Si platform, researchers are harnessing different material platforms to overcome the limitations associated with Si. 

Over the years, stoichiometric silicon nitride $(Si_{3}N_{4})$ material has gained significant attention as a promising CMOS platform for building photonic integrated circuits (PICs) \cite{moss2013new} . Apart from its CMOS compatibility, it offers multiple advantages, such as a very large band gap with low propagation loss, high-refractive index contrast with silicon oxide $(SiO_{2})$; absence of TPA in the telecom wavelength range; and it offers relatively low TOC \((dn/dt = 2.45 \times 10^{-5} K^{-1})\) compared to both crystalline and amorphous Si \cite{arbabi2013measurements}. These properties of $(Si_{3}N_{4})$ led to employ it in a multitude of applications ranging from nonlinear photonics \cite{pfeiffer2017octave}, high Q resonator \cite{luke2015broadband}, and all-optical signal processing \cite{sharma2020review}. However, despite many conveniences, $Si_{3}N_{4}$ suffers from potential drawbacks, mainly due to the low refractive index contrast between core and cladding layers in the SOI platform, which in turn increases the device footprint and also low Kerr nonlinearity. 

However, these bottlenecks have been effectively addressed by tuning the optical properties of SiN through proper control over the N/Si ratio of the material \cite{tan2018nonlinear}. Si-rich SiN (SRSN) material has attracted considerable interest due to its fabrication flexibility and the ability to tailor the intrinsic properties between those of $Si_{3}N_{4}$ and Si \cite{sohn2019optical}. By adjusting the ratio of the precursor gases (\(SiH_{4}/N_{2}\)) in the plasma-enhanced chemical vapor deposition (PECVD) process, the percentage of Si content in the material can be effectively tuned which provides effective route to engineer the optical properties of SRSN material \cite{tan2021nonlinear}. Proper N/Si ratio exhibits optical band gap variation from 2.7 to 5 eV and refractive index (n) varies over a wide range from 1.91 to 3.1 at 1550 nm wavelength \cite{nejadriahi2020thermo}. On the other hand, the third-order nonlinear coefficient increases effectively with the increase in Si content in the material \cite{kruckel2015linear}, which has already been demonstrated in a variety of applications such as parametric gain, broadband supercontinuum generation \cite{liu2016octave}, nonlinear signal processing \cite{dizaji2017silicon, lacava2017si}, and wavelength conversion \cite{yang2018octave,lacava2019intermodal}. The TOC as well as linear and nonlinear refractive index of SRSN material also increases linearly with the increase in the fractional composition of silicon over a range from that of silicon nitride to a-Si \cite{pruiti2020thermo,nejadriahi2020thermo}. Hence, careful control over the Si composition in the material is required to realize a thermally stable and high refractive index contrast integrated CMOS SRN platform.

\begin{figure}[t]
\centering
{\includegraphics[width=1\linewidth,height=4.5 cm]{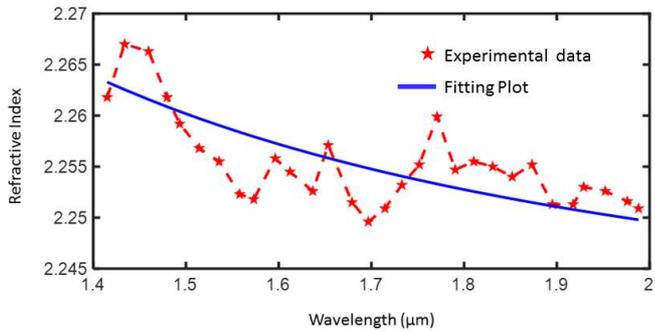}}
\caption{Refractive index as measured by ellipsometry for the SRN film.}
\label{fig:refractive_index_label}
\end{figure}

\begin{figure}[t]
\centering
   \begin{subfigure}[]{0.4\textwidth}
 {\includegraphics[width=1\linewidth ,height=4 cm]{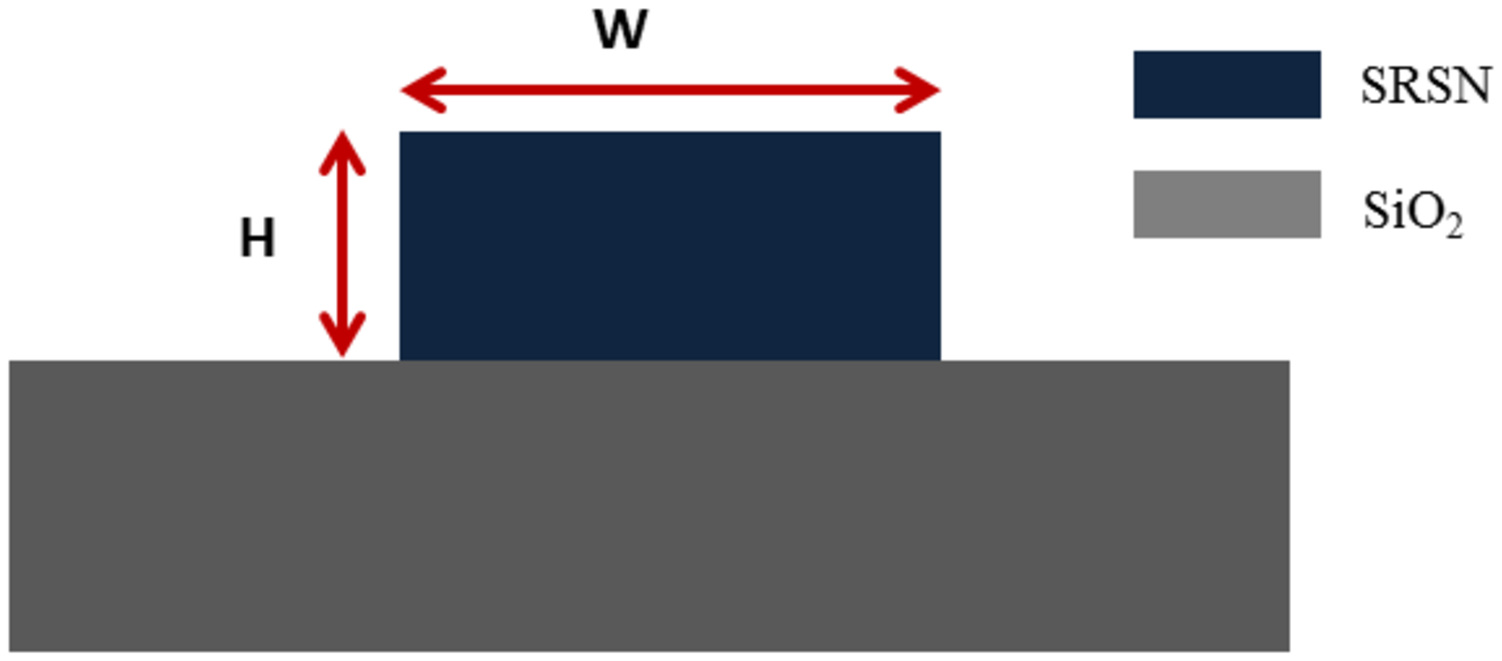}}
   \caption{}
   \label{fig:2d_diagram_label} 
\end{subfigure}
\begin{subfigure}[]{0.4\textwidth}
 {\includegraphics[width=0.9\linewidth ,height=3.5 cm]{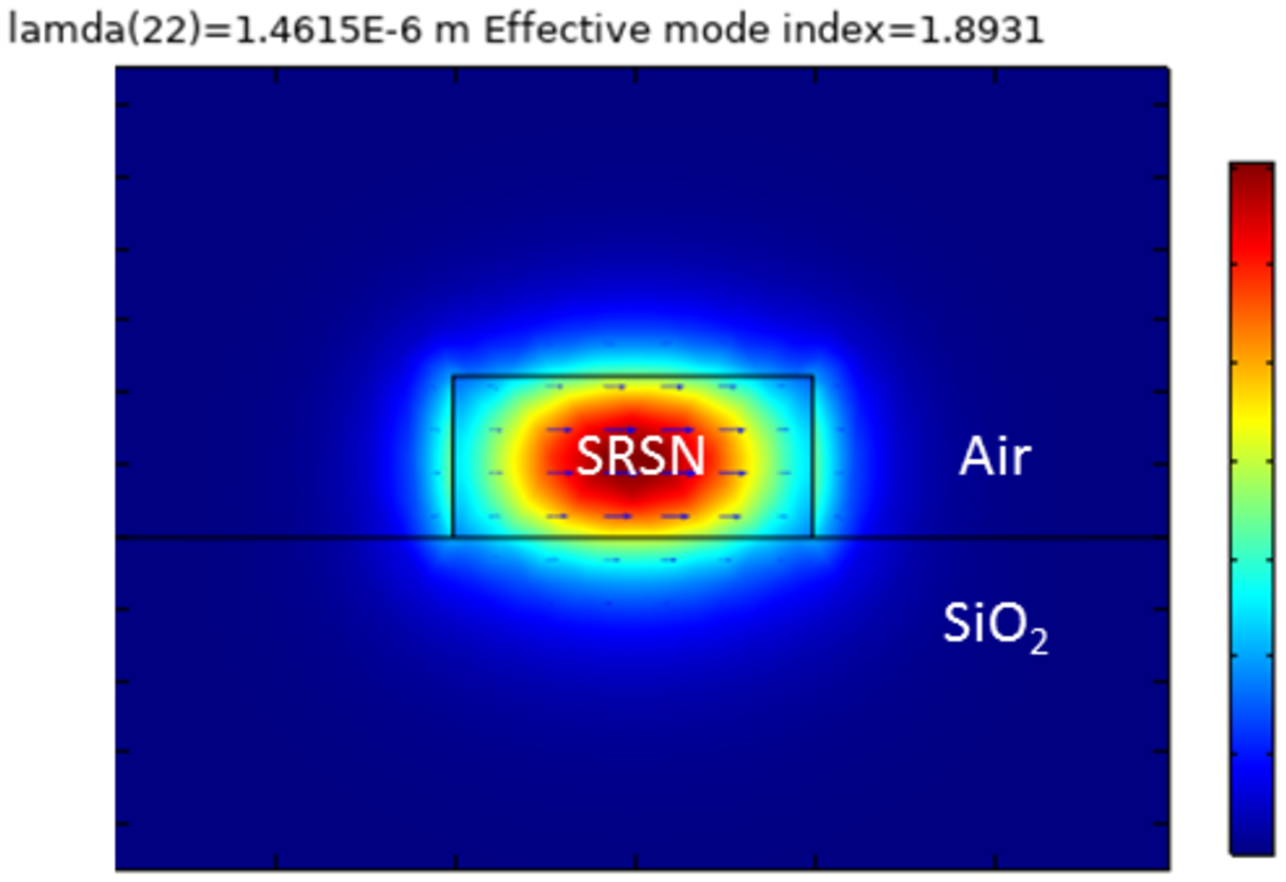}}
   \caption{}
   \label{fig:mode_profile_label}
\end{subfigure}
\begin{subfigure}[]{0.4\textwidth}
\centering
 {\includegraphics[width=1\linewidth ,height=4 cm]{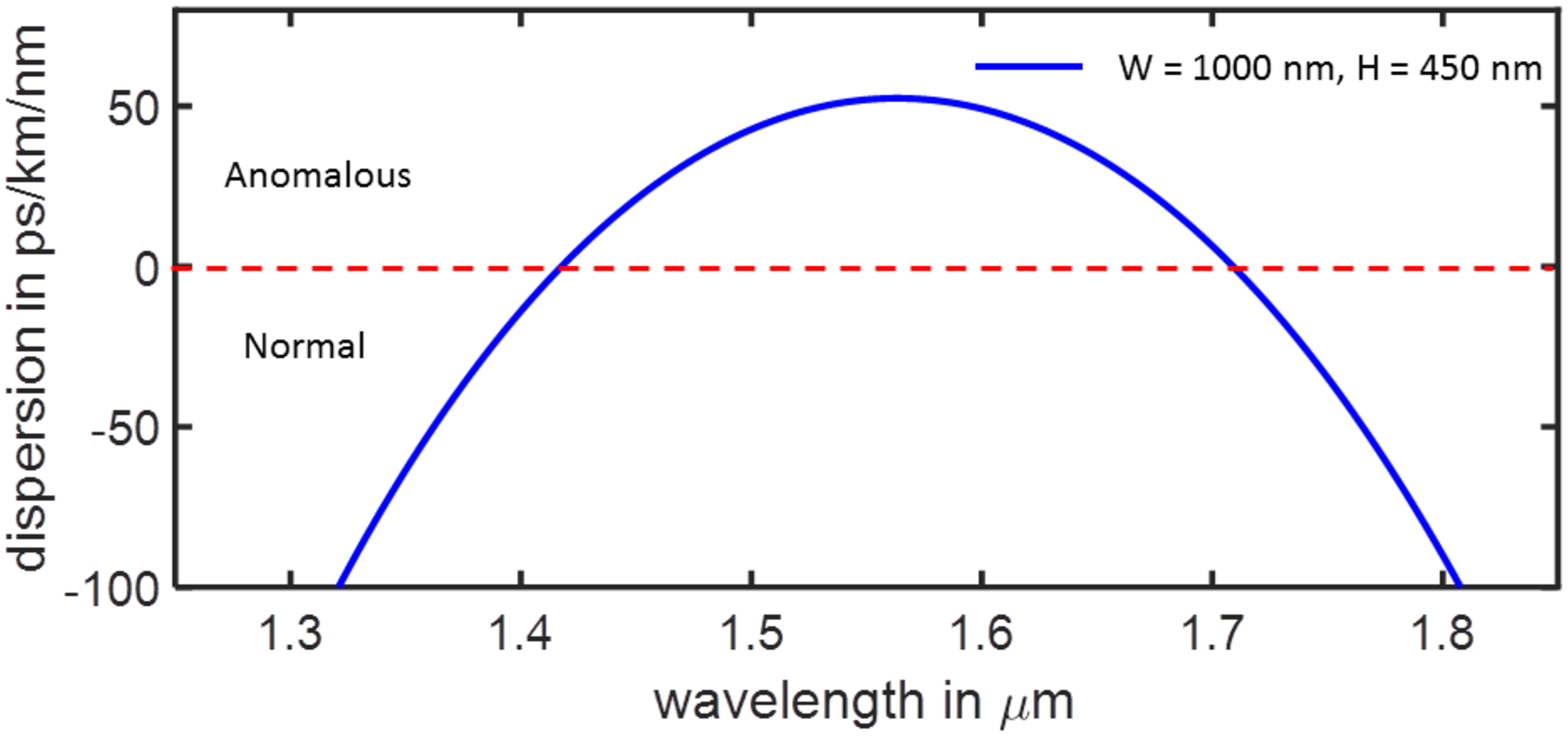}}
   \caption{}
   \label{fig:dispersion_label}
\end{subfigure}

\caption{(a) Schematic of the transverse view of the SRN waveguide (b) Modal profile of the fundamental TE mode of the proposed waveguide (c) Dispersion plot of the SRN waveguide structure. }
\end{figure}

In this manuscript, we demonstrate a broadband dispersion engineered racetrack ring resonator structure fabricated with the SRN material platform to realize on-chip applications in the telecommunication band. We designed and fabricated grating couplers at the input and output terminals to couple the light into the device and perform optical characterizations. We measure the TOC of PECVD deposited SRSN that exhibits a refractive index of 2.25 at 1550 nm. The TOC of the SRSN waveguide has been measured by determining the temperature-dependent resonance shift of the racetrack ring resonator structure in the transmission spectrum. Furthermore, we investigate power-dependent transmission spectra to characterize the thermal stability and nonlinear loss in the communication wavelength range. 
\section{Design and fabrication of SRSN Waveguide}

\begin{figure}[t]
\centering
{\includegraphics[width=1\linewidth,height=4.5 cm]{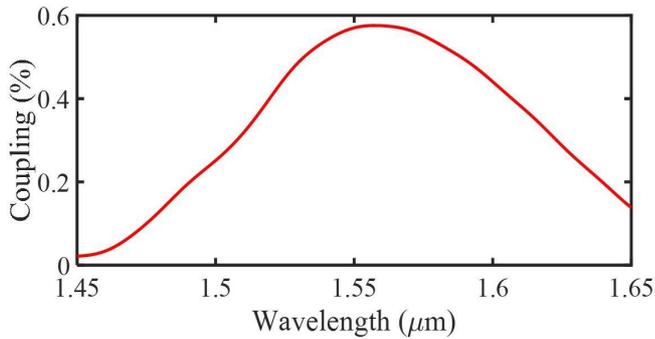}}
\caption{Coupling efficiency of the grating coupler simulated through Lumerical FDTD software.}
\label{fig:coupling_efficiency}
\end{figure}

\begin{figure}[t]
\centering
\begin{subfigure}[]{0.4\textwidth}
 {\includegraphics[width=1.1\linewidth ,height=3.5 cm]{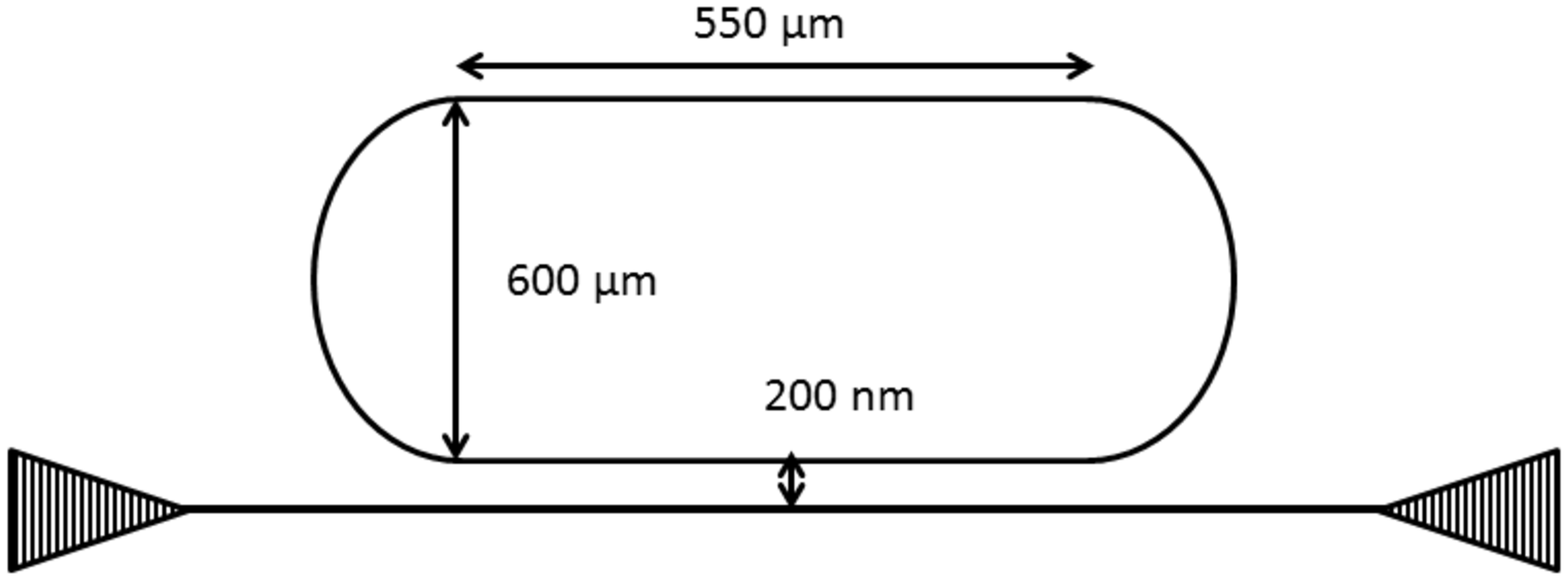}}
   \caption{}
   \label{fig:racetrack_ring_label}
\end{subfigure}
   \begin{subfigure}[]{0.4\textwidth}
 {\includegraphics[width=1.1\linewidth ,height=3.5 cm]{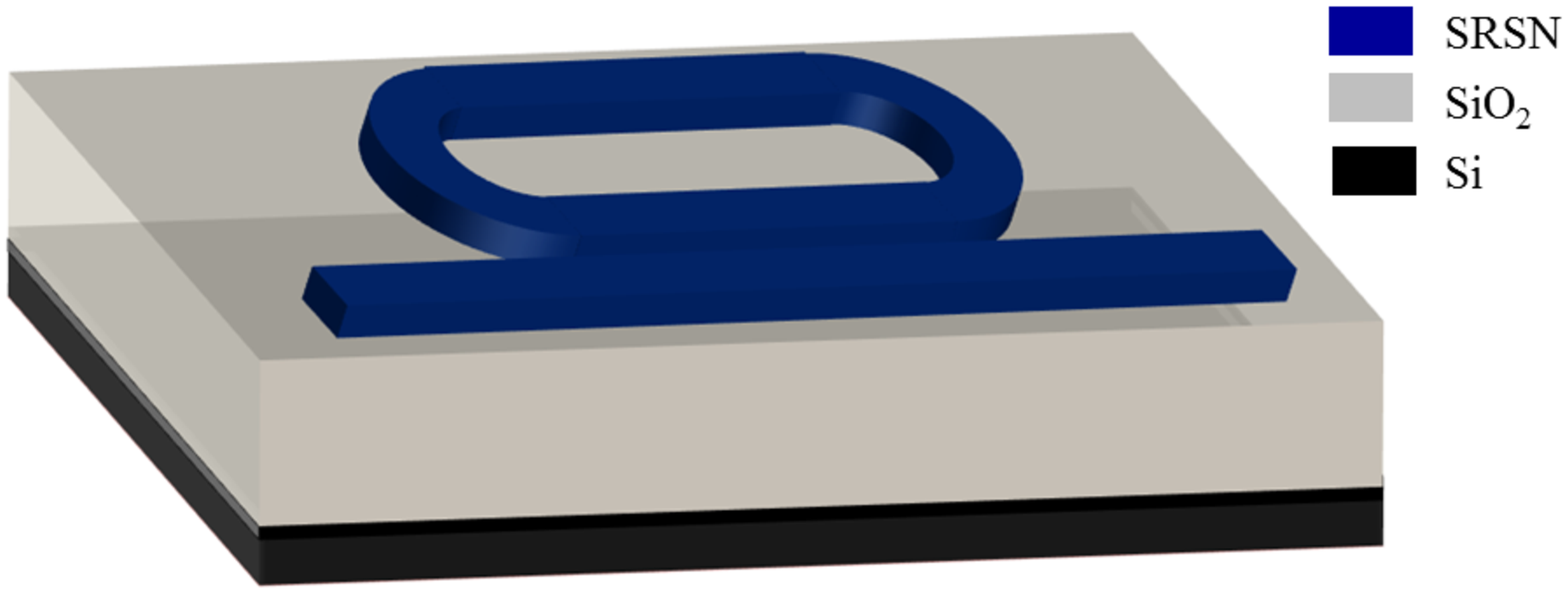}}
   \caption{}
   \label{fig:racetrack_ring_3d_label} 
\end{subfigure}
\begin{subfigure}[]{0.4\textwidth}
 {\includegraphics[width=1\linewidth ,height=3.6 cm]{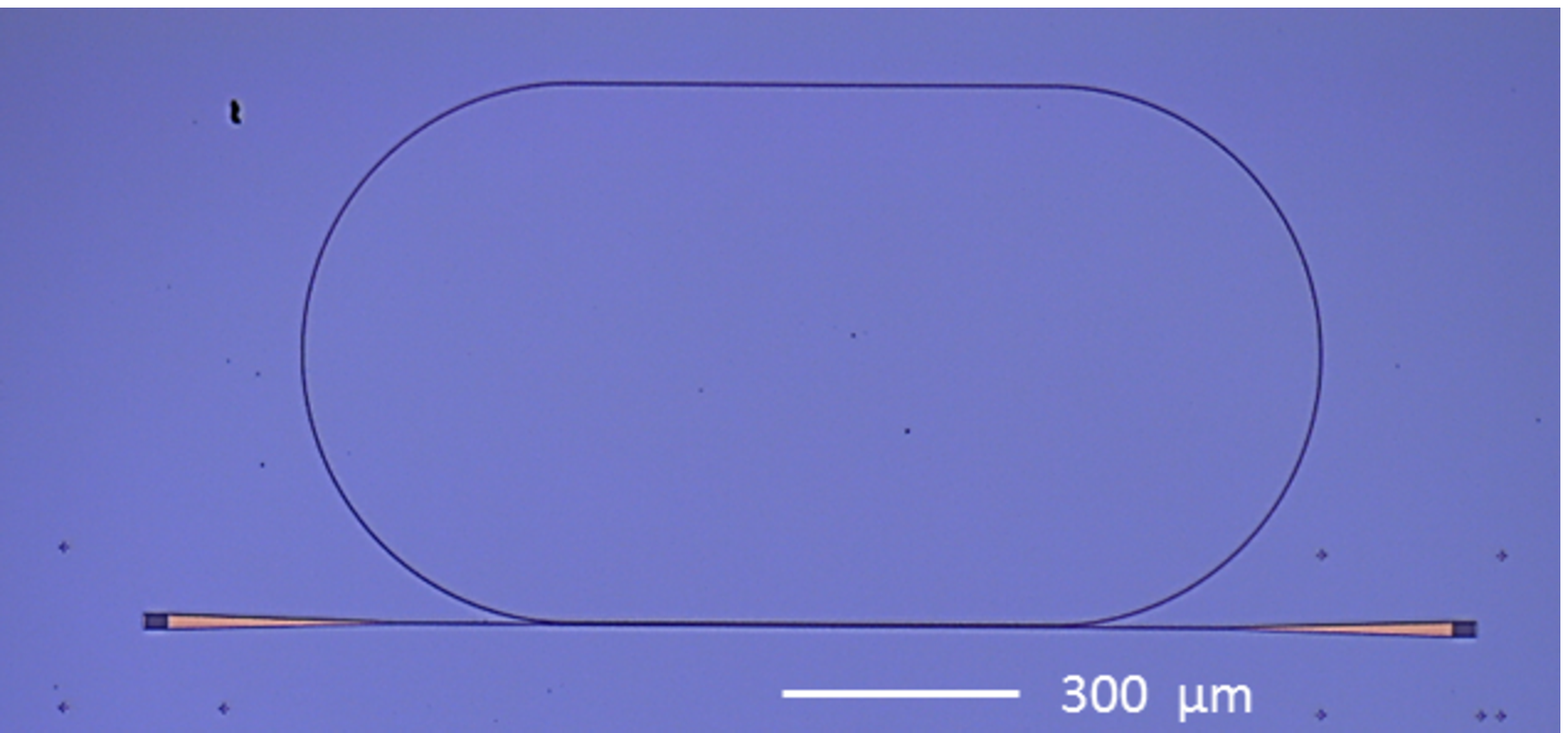}}
   \caption{}
   \label{fig:racetrack_ring_microscope_label}
\end{subfigure}
\caption{(a) Schematic diagram of the racetrack ring resonator structure. (b) Perspective view of the racetrack ring resonator and the bus waveguide. (c) Optical microscopy image of the racetrack ring resonator along with the bus waveguide and the grating coupler.}
\end{figure}

For fabrication, we have employed the PECVD deposition technique to deposit SRSN on the top of the Si wafer with a 3 \(\mu\)m buried oxide layer. The deposition was done at a low temperature (\(350^{0}\) C) and comprises precursor gases like \(NH_{3}, SiH_{4}\) , and \(N_{2}\) with different flow rates. To characterize the real and imaginary parts of the refractive index of SRSN film, we deposit 100 nm of SRSN on the top of a clean Si wafer. After that, we carried out ellipsometry measurements with a broad range of wavelengths. Fig. \ref{fig:refractive_index_label} shows the dispersion of the real and imaginary parts of the refractive index of the deposited SRSN layer. It is observed that the value of the real part of the refractive index is significantly higher than that of silicon nitride material. We have employed this refractive index profile to design our SRSN-based racetrack ring resonator structure. 

The schematic of the waveguide structure with optimized waveguide parameters is shown in Fig. \ref{fig:2d_diagram_label}. The waveguide parameters have been optimized using full-vectorial mode solver COMSOL software to achieve a flat dispersion profile in the telecom wavelength range.  The optimized height (H) and width (W) for the SRSN waveguide are 450 nm and 1000 nm, respectively. Fig. \ref{fig:mode_profile_label} shows the modal confinement of the fundamental TE mode at 1550 nm for the optimized waveguide parameters. Whereas the dispersion profile is shown in Fig. \ref{fig:dispersion_label}, which exhibits flat dispersion in the telecom band and can be useful for complex nonlinear applications. 

\begin{figure*}[htp]
\centering
{\includegraphics[width=0.8\linewidth,height=8 cm]{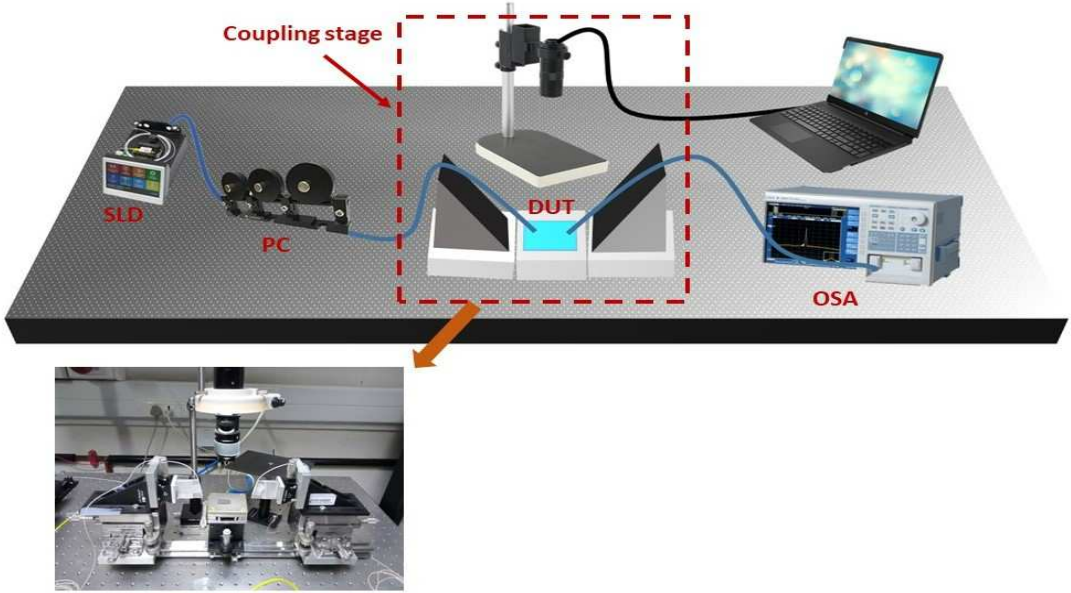}}
\caption{Schematic of experimental setup for characterization, SLD: Superluminescent Diode source, PC: polarization controller, DUT: Device under test, OSA: Optical spectrum analyzer.}
\label{fig:exp_setup}
 \end{figure*}
 
The optimized waveguide parameters have been employed for fabricating the racetrack ring resonator structure. The schematic of the racetrack ring resonator structure is shown in Fig. \ref{fig:racetrack_ring_label}. The racetrack resonator is designed for operating in the telecommunication wavelength range. The gap and coupling length have been chosen as 200 nm and 550 \(\mu\)m. To achieve maximum coupling, the grating coupler has been designed with an SRSN width of 285 nm, a shallow-etched depth of 250 nm, and a grating period of 870 nm. The coupling efficiency of the optimized design is shown in Fig. \ref{fig:coupling_efficiency}, which is simulated through Lumerical FDTD software. The 3-dB bandwidth of the grating coupler is calculated as 50 nm. To fabricate the device, we deposit a 3 \(\mu\)m \(SiO_{2}\) buried oxide layer on the top of a clean Si wafer by the PECVD method. After that, a 450 nm thick SRSN layer is deposited at low temperature using the PECVD method. After depositing the layers, the racetrack ring and the grating coupler are patterned employing electron-beam lithography (Raith-Eline) with a negative MaN 2401 resist, followed by reactive ion etching (RIE), with fluorine chemistry. Fig. \ref{fig:racetrack_ring_microscope_label} shows the microscopic view of the fabricated device. 
\begin{figure}[htp]
\centering
{\includegraphics[width=1\linewidth,height=6 cm]{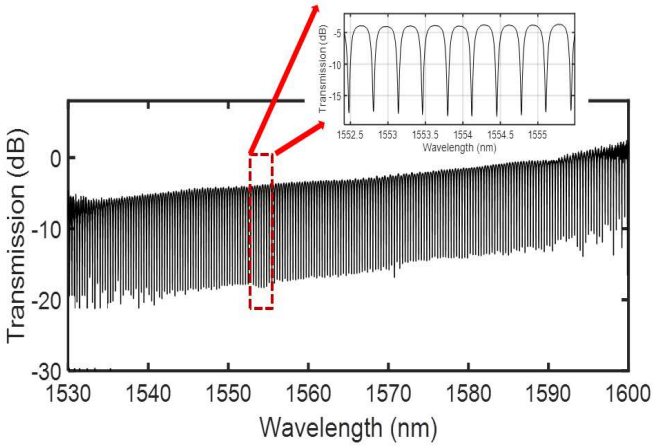}}
\caption{The measured transmission spectrum of the racetrack ring resonator at the through port for the TE mode.}
\label{fig:transmission}
\end{figure}

Device characterization is performed using a TEC-enabled broadband superluminescent compact laser diode (Thorlabs 1550S-A2), which is connected to the input probe with a single-mode fiber passing through a polarization controller and three-axis translation stage. The fiber incident angle is set at \(10 ^{0}\). The output signal was collected with an optical spectrum analyzer (Yokogawa
AQ6370D) ranging from 600 to 1700 nm. The schematic of the experimental setup is shown in Fig. \ref{fig:exp_setup}. The photograph of the actual coupling stage is shown on the inset of the diagram. The normalized TE transmission is shown in Fig. \ref{fig:transmission}. The measured spectrum shows periodic resonance dips with uniform spacing between the adjacent dips. The quality factor (Q) measured at 1553.797 nm is \(22.5 \times 10^{3}\), whereas the FWHM linewidth and free spectral range (FSR) are measured as 0.069 nm and 0.326 nm, respectively.

\section{Thermo-optic characterization}

To characterize the thermal stability of the device, we have performed TOC measurements of the SRSN platform by calculating the wavelength shift \(\Delta \lambda\) of the resonant peak with the change in the environmental temperature. The wavelength shift can be calculated as
\begin{equation}
\Delta\lambda = \bigg(\dfrac{\partial n_{eff}}{\partial T_{R}} + n_{eff}. \alpha_{sub}\bigg) \dfrac{\lambda}{n_{g}} T_{R} 
 \label{eq1}
\end{equation}
where \(\lambda\) is the resonance wavelength at room temperature, \(\alpha_{sub}\) is the thermal expansion coefficient of the substrate, and \(\Delta T_{R}\) is the change in environmental temperature. \(n_{g}\) is the group index which can be expressed as \(n_{g} = \dfrac{\lambda^{2}}{L_{R}.FSR}\), \(L_{R}\) and FSR are the resonator length and free-spectral range, respectively. The effective TOC defines the change in the effective index as a function of temperature, which can be expressed as   
\begin{equation}
\kappa_{eff} = \dfrac{\partial n_{eff}}{\partial T_{R}} = \bigg(\dfrac{n_{g}}{\lambda} \dfrac{\partial\lambda}{\partial T_{R}} - n_{eff}.\alpha_{sub} \bigg)
 \label{eq2}
\end{equation}
where \(\dfrac{\partial\lambda}{\partial T_{R}}\) defines the wavelength shift of the resonance dip as a function of temperature. 
However, as the effective TOC involves the modal overlapping between the core of the SRSN waveguide and cladding oxide, eq. \ref{eq2} can be rewritten considering the overlapping factor as
\begin{equation}
\kappa_{eff} =  \Gamma_{SiN_{x}} \bigg(\dfrac{\partial n}{\partial T} \bigg)_{SiN_{x}} + \Gamma_{SiO_{2}} \bigg(\dfrac{\partial n}{\partial T} \bigg)_{SiO_{2}}
 \label{eq3}
\end{equation}
where \(\Gamma_{SiN_{x}}\) and \(\Gamma_{SiO_{2}}\) are the modal overlapping factor with SRSN waveguide core and cladding oxide, respectively.
\begin{figure}[htp]
\centering
   \begin{subfigure}[]{0.4\textwidth}
 {\includegraphics[width=1\linewidth ,height=3.5 cm]{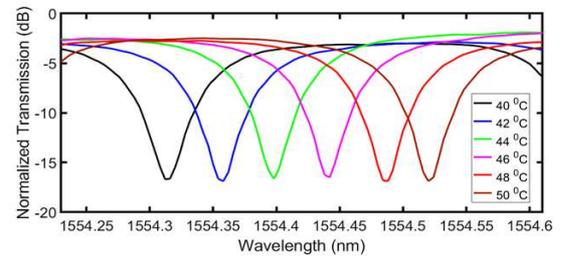}}
   \caption{}
   \label{fig:TOC_shift_label} 
\end{subfigure}
\begin{subfigure}[]{0.4\textwidth}
 {\includegraphics[width=1\linewidth ,height=3.1 cm]{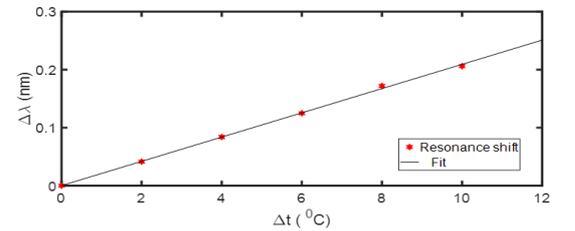}}
   \caption{}
   \label{fig:TOC_shift_temp_label}
\end{subfigure}
\caption{(a) Spectral response of the resonance dip at different temperature . (b) Plot of the resonance shift \(\Delta\lambda\) as a function of substrate temperature.}
\end{figure}

\begin{figure}[htp]
\centering
   \begin{subfigure}[]{0.48\textwidth}
 {\includegraphics[width=1\linewidth ,height=4 cm]{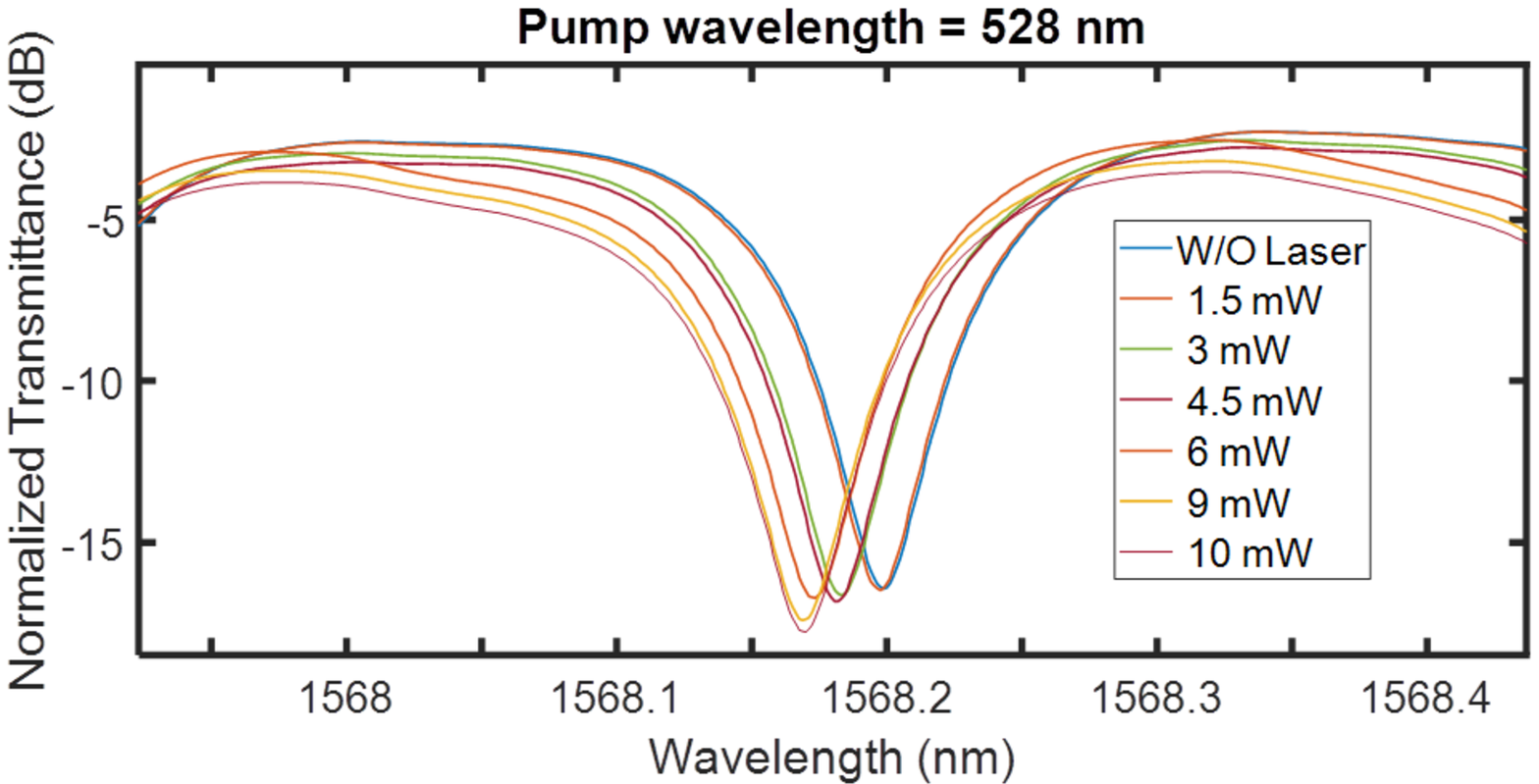}}
   \caption{}
   \label{fig:TPA_528_label} 
\end{subfigure}
\begin{subfigure}[]{0.48\textwidth}
 {\includegraphics[width=1\linewidth ,height=3.7 cm]{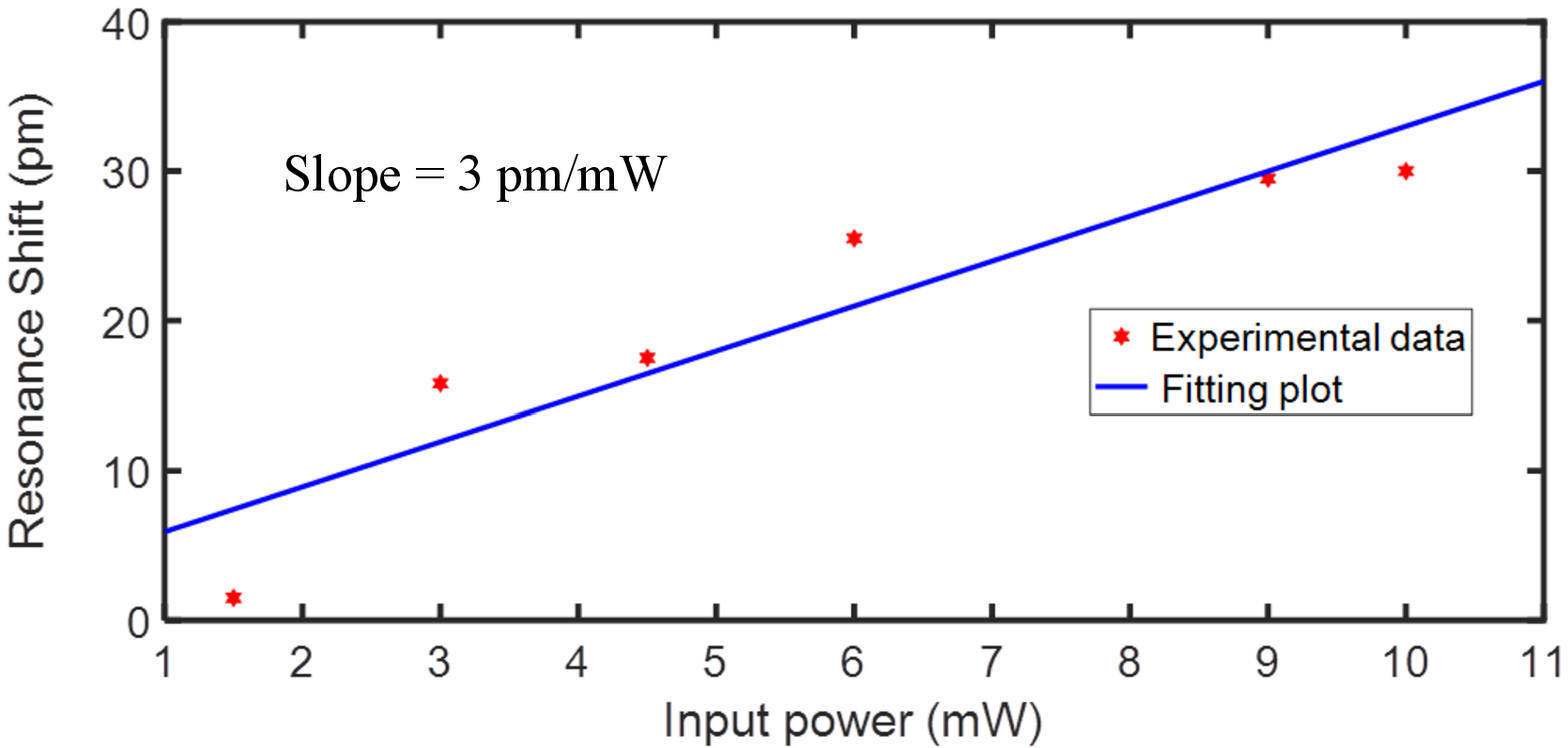}}
   \caption{}
   \label{fig:temp_shift_528nm_label}
\end{subfigure}
\caption{(a) Power-dependent transmission spectra for diode pump laser at 528 nm (b) corresponding spectral shift of resonance wavelength with the pump power variation.}
\end{figure}

\begin{figure}[htp]
\centering
   \begin{subfigure}[]{0.45\textwidth}
 {\includegraphics[width=1\linewidth ,height=3.8 cm]{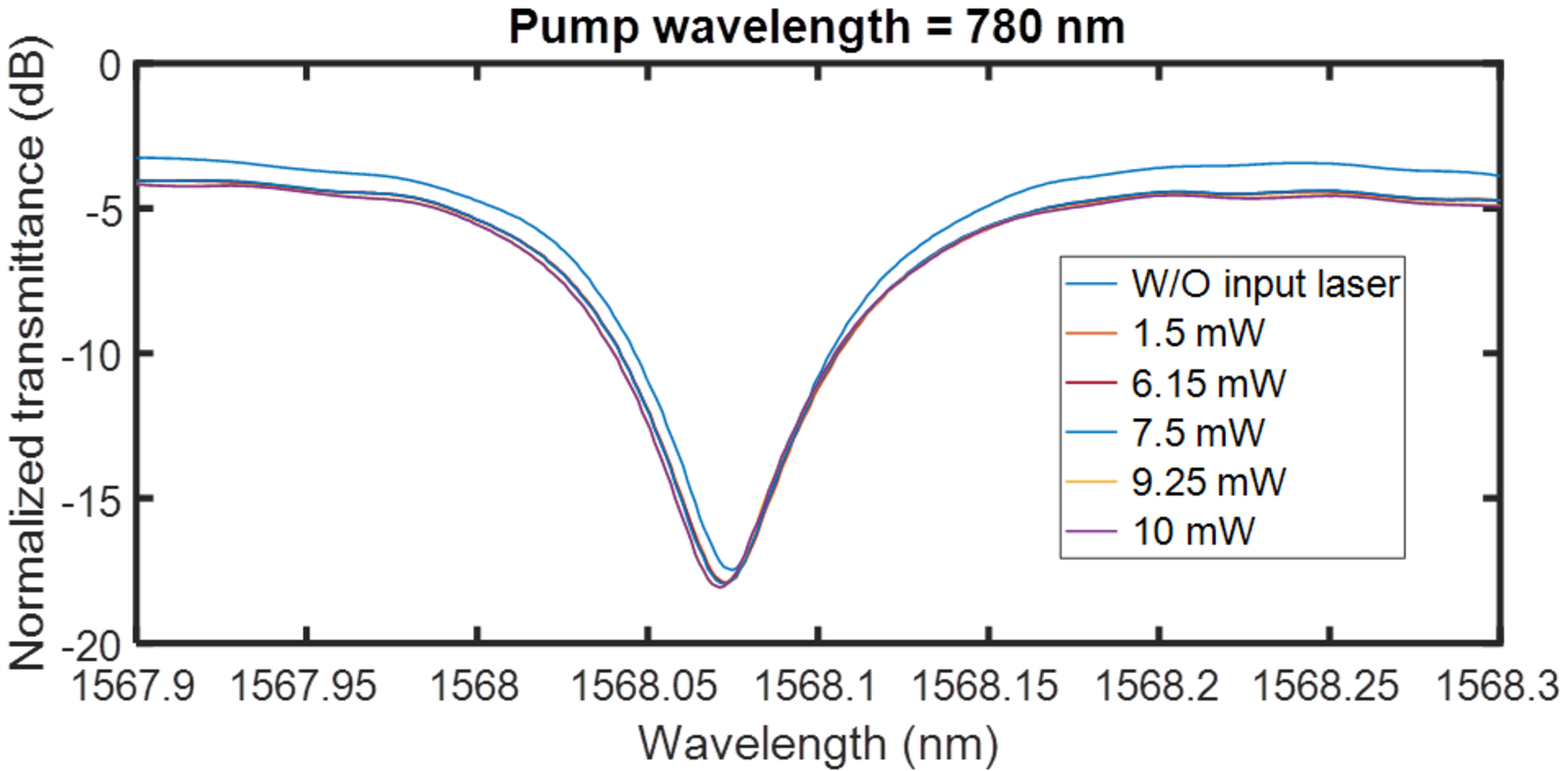}}
   \caption{}
   \label{fig:TPA_780_label} 
\end{subfigure}
\begin{subfigure}[]{0.45\textwidth}
 {\includegraphics[width=1\linewidth ,height=3.5 cm]{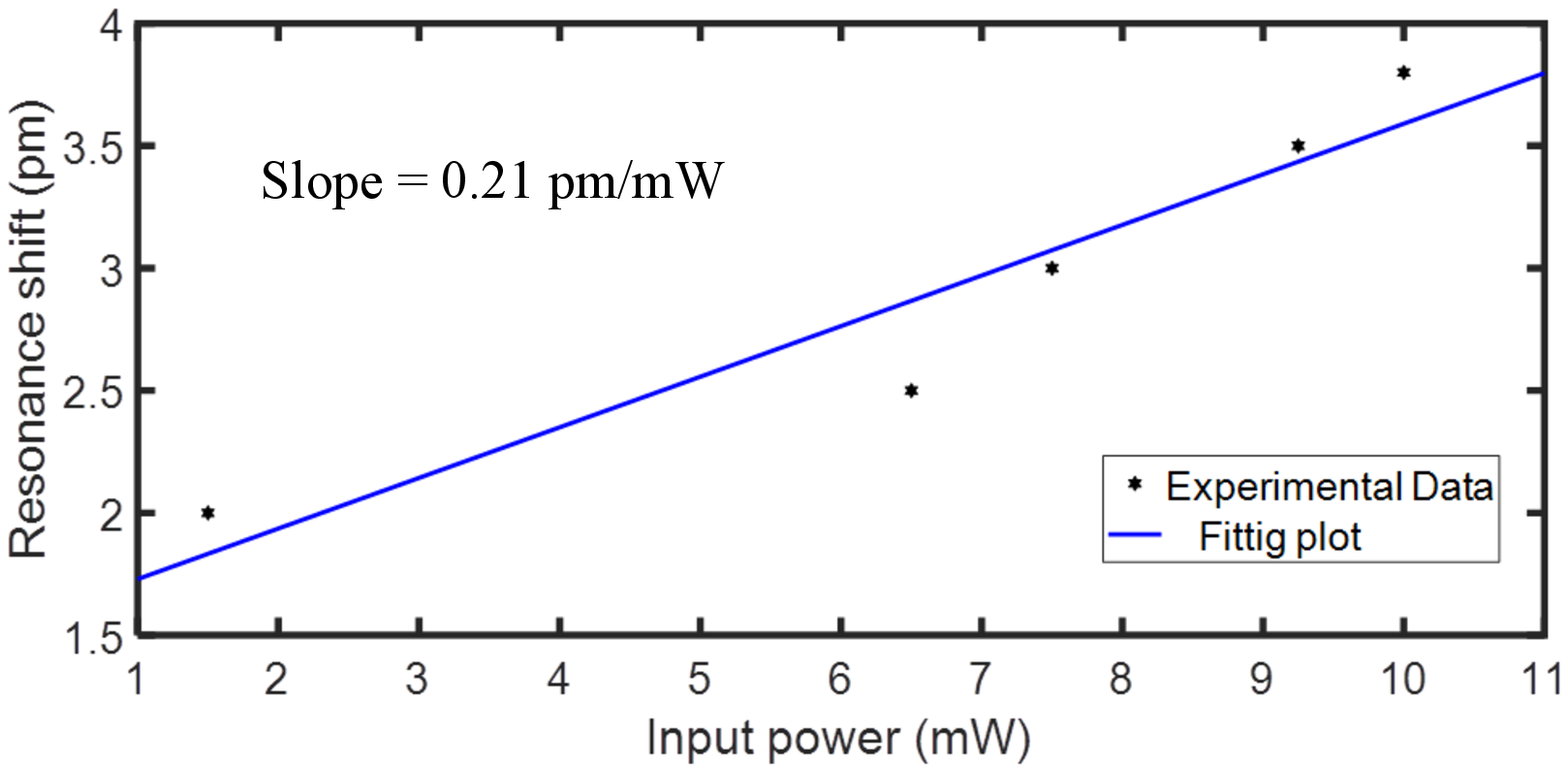}}
   \caption{}
   \label{fig:temp_shift_780nm_label}
\end{subfigure}
\caption{(a) Power-dependent transmission spectra for diode pump laser at 780 nm (b) corresponding spectral shift of resonance wavelength with the pump power variation.}
\end{figure}

\begin{figure}[htp]
\centering
   \begin{subfigure}[]{0.45\textwidth}
 {\includegraphics[width=1\linewidth ,height=3.8 cm]{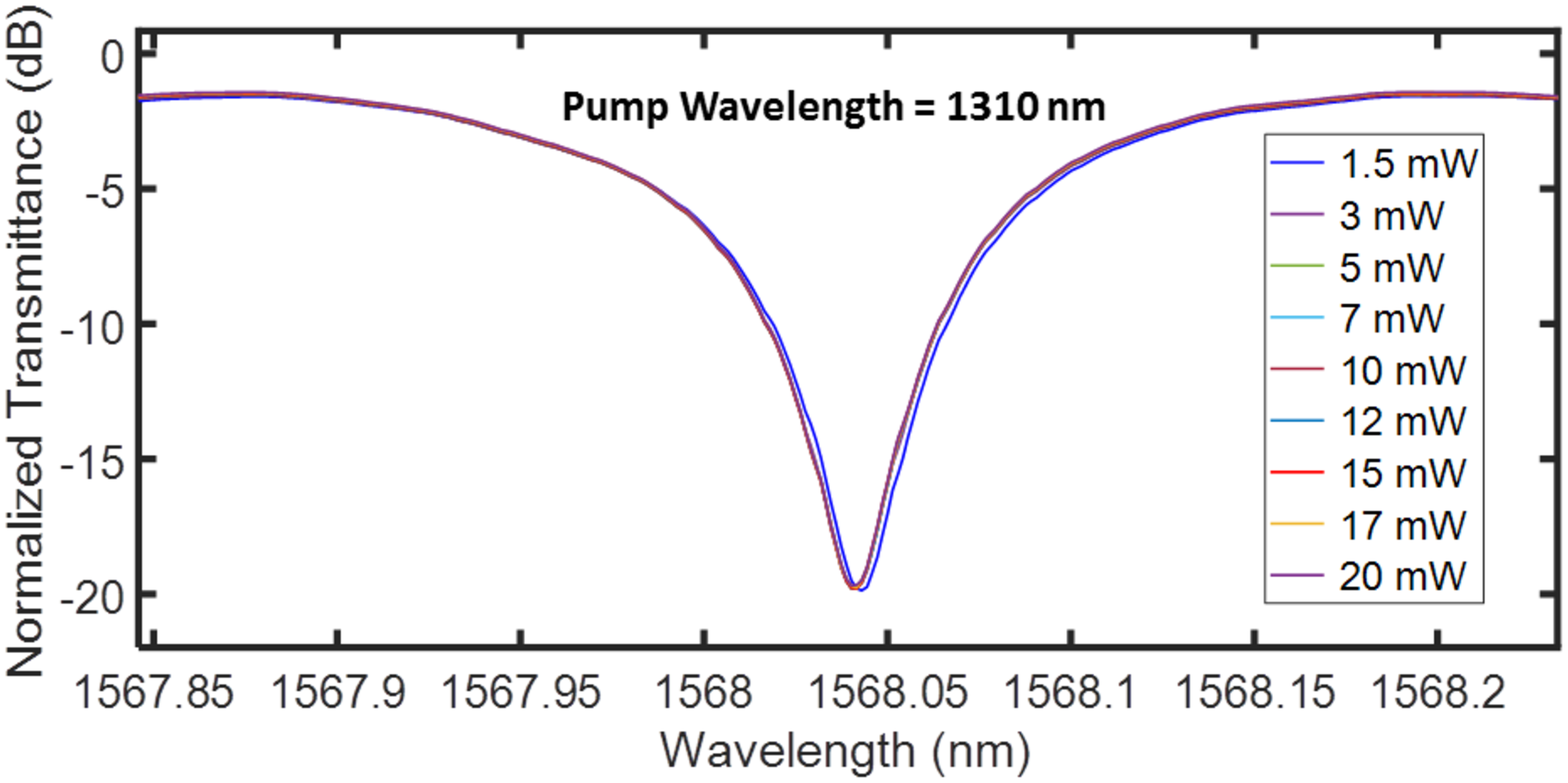}}
   \caption{}
   \label{fig:TPA_1310_label} 
\end{subfigure}
\begin{subfigure}[]{0.45\textwidth}
 {\includegraphics[width=1\linewidth ,height=3.8 cm]{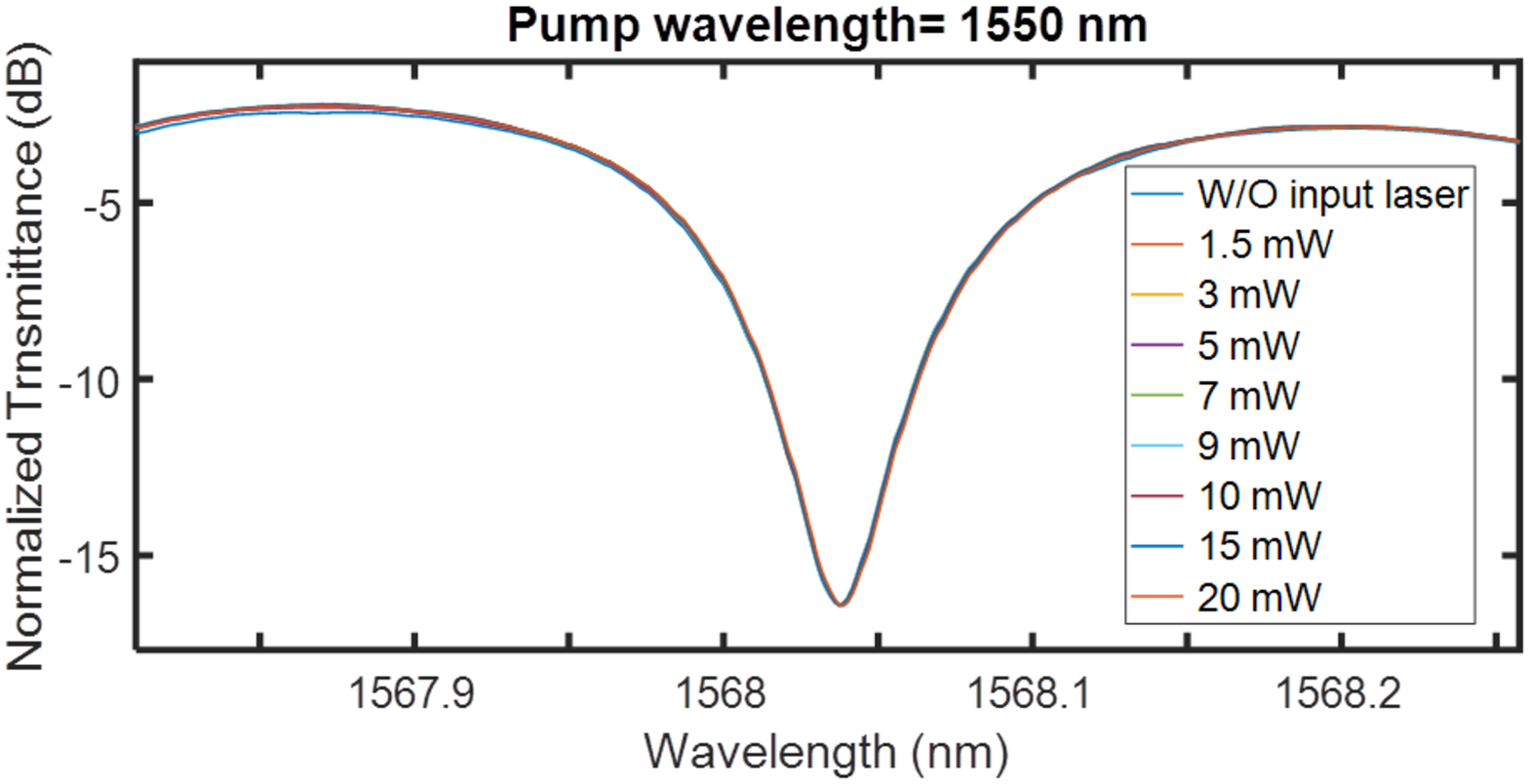}}
   \caption{}
   \label{fig:TPA_1550_label}
\end{subfigure}
\caption{Power-dependent transmission spectra for diode pump laser at (a) 1310 nm (b) 1550 nm wavelength.}
\end{figure}

To extract the TOC of the SRSN waveguide, we performed a temperature-dependent measurement of the racetrack ring resonator. The temperature-dependent transmission spectrum for the resonator structure is shown in Fig. \ref{fig:TOC_shift_label}, where the substrate temperature was varied across the range between \(40^{0}\)C to \(50^{0}\)C. The resonance dip at \(\sim\) 1554.32 nm exhibits red-shifting with a value of \(\sim\) 200 pm when the substrate temperature is increased from \(40^{0}\)C to \(50^{0}\)C. The \({\partial\lambda}/{\partial T_{R}}\) value has been calculated by performing linear fitting of the resonator wavelength shift against the substrate temperature as shown in Fig. \ref{fig:TOC_shift_temp_label} and the value obtained as 0.021 nm/\(^{0}C\). Taking \(L_{R}\) and FSR values as 2984 \(\mu\)m and 0.326 nm, respectively, the group index \(n_{g}\) value is calculated as 2.4829.  The calculated values of the overlapping factor for \(\Gamma_{SiN_{x}}\), \(\Gamma_{SiO_{2}}\) , effective mode indices \(n_{eff}\) are given by 0.846 and 0.096, and 1.854, respectively, which were obtained through the finite difference method. Furthermore, for the calculation, we consider \(\alpha_{sub}\) = \(2.6 \times 10^{-6}\). Putting all these values in eq. \ref{eq2}, we calculate the \(\kappa_{eff}\) as \(2.873\) $\times$ \(10^{-3}\) \(^{0} C^{-1}\). Considering \((\partial n/ \partial T)_{SiO_{2}}\) = 1 $\times$ \(10^{-5}\)  \(^{o} C^{-1}\), TOC of the SRSN waveguide is determined from eq. \ref{eq3} to be \(3.2825\) $\times$ \(10^{-5}\)\(^{o} C^{-1}\). It can be seen that the measured value of TOC is well below that of Si.

\section{TPA characterization}

Additionally, in order to characterize the thermal stability and nonlinear losses in the telecommunication band, we measure the power-dependent transmission spectrum. Optical characterization was performed using a tunable laser source coupled with a diode laser source at 528 nm, 780 nm, 1310 nm, and 1550 nm wavelengths. Finally, the output signal was collected by OSA. The transmission spectrum is measured as a function of input optical power. The power-dependent transmission spectrum for the diode laser pumped at 528 nm is shown in Fig. \ref{fig:TPA_528_label}, where a blue-shifted spectral response with an increase in input pump power is observed. The corresponding spectral shift with the incident pump power is shown in Fig. \ref{fig:temp_shift_528nm_label}. The fitting line shows a spectral shift of 3 pm/mW. It is important to note that the direction of the wavelength shifting depends on the dominant nonlinear process. The incident light gets absorbed via TPA, which in turn generates free carriers. The generated free carriers cause blue-shifting of the resonance peaks owing to free carrier dispersion (FCD). On the other hand, the generated free carriers trigger the free carrier absorption (FCA) loss. The combined effect of TPA, FCA, and the linear surface absorption leads to heating of the resonator, where the thermo-optic effect causes red-shifting of the resonance peak \cite{luo2012power,gao2017cavity}. As Si has a high positive TOC, thermal effects dominate and exhibit red-shifting in Si-based devices \cite{jeyaselvan2021mitigation, xiang2020effects}. In contrast, in our case, we observe a blue-shifting of the resonance peak. This is because the SRSN has a much lower TOC as measured in the previous section compared to Si. Hence, the FCD mechanism dominates over the thermo-optic effect, resulting in blue-shifting of the spectral peaks. Fig. \ref{fig:TPA_780_label} exhibits power-dependent transmission spectra for the pump wavelength at 780 nm. A small spectral blue shift is observed at that wavelength. The variation of the resonance shift with the input pump power is shown in Fig. \ref{fig:temp_shift_780nm_label}. The fitting line of the resonance shift exhibits a spectral shift of 0.21 pm/ mW at the wavelength of 780 nm wavelength, which is much lower than that observed at 580 nm. This is because the TPA effect gets reduced at 780 nm wavelength compared to 528 nm. The power-dependent transmission spectrum for a diode laser pumped at 1310 nm is shown in Fig. \ref{fig:TPA_1310_label}. The input power of the diode laser varied from 1.5 mW to 20 mW. It is observed that the SRSN-based device exhibits efficient thermal stability and suffers from no nonlinear losses as the resonance shape remains almost unchanged. On the other hand, the power dependent-transmission spectrum for diode laser at 1550 nm is shown in Fig. \ref{fig:TPA_1550_label}, which also exhibits remarkable thermal stability and the absence of nonlinear losses. Therefore, our results indicate that the SRSN device offers extensive thermal stability and the nonexistence of nonlinear losses in the telecommunication band.

\section{Conclusion}

In conclusion, we have demonstrated an SRSN-based broadband integrated racetrack ring resonator for on-chip applications in the telecommunication wavelength range. SRSN material exhibits an enhanced refractive index of 2.25 at 1550 nm. The TOC of the SRSN platform is characterized by measuring the temperature-dependent transmission spectra for the resonator structure, and the TOC is estimated to be \(3.2825\) $\times$ \(10^{-5}\)\(^{o}\) \(C^{-1}\). Furthermore, we have measured the power-dependent transmission spectra in the communication band. Transmission spectra almost remain unchanged for high input power, which reveals significant thermal stability, and the absence of nonlinear absorption. Our experimental findings pave the way toward the implementation of on-chip devices for linear and nonlinear applications in the telecommunication band.  
\section*{Acknowledgments}

DST/Nano Mission and Ministry of Electronics and Information Technology.



\bibliography{references}

\end{document}